\def\Pb{P_0}

\def\rfr#1{eq. (\ref{#1})}

\def\dert#1#2{\frac{{{d}}{#1}}{{{d}}{#2}}}

\def\eqi{\begin{equation}}
\def\eqf{\end{equation}}
\def\eqia{\begin{eqnarray}}
\def\eqfa{\end{eqnarray}}

\def\rp#1#2{{#1\over#2}}
\def\lb#1{\label{#1}}

\def\bds#1{\boldsymbol{#1}}

\def\ton#1{\left(#1\right)}
\def\qua#1{\left[#1\right]}
\def\grf#1{\left\{#1\right\}}
\def\ang#1{\left\langle #1\right\rangle}

\documentclass{aastex}
\usepackage{amsmath,textgreek,w-greek,wasysym}
\usepackage{amscd,lineno}
\usepackage{amssymb}
\usepackage{graphicx,epsfig}
\usepackage{txfonts}
\RequirePackage{color}

\newcommand{\emaila}{lorenzo.iorio@libero.it}

\allowdisplaybreaks[1]

\begin{document}

\title{Does the Newton's gravitational constant vary sinusoidally with time? Orbital motions say no}

\shortauthors{L. Iorio}

\author{Lorenzo Iorio\altaffilmark{1} }
\affil{Ministero dell'Istruzione, dell'Universit\`{a} e della Ricerca (M.I.U.R.)-Istruzione
\\ Permanent address for correspondence: Viale Unit\`{a} di Italia 68, 70125, Bari (BA), Italy.}

\email{\emaila}

\begin{abstract}
A sinusoidally time-varying pattern of the values of the Newton's constant of gravitation $G$ measured in Earth-based laboratories over the latest decades has been recently reported in the literature.
We put to the test the hypothesis that the aforementioned harmonic variation  may pertain $G$ itself in a direct and independent way.
We numerically integrated the ad-hoc modified equations of motion of the major bodies of the Solar System by finding that the orbits of the planets would be altered by an unacceptably larger amount in view of the present-day high accuracy astrometric measurements. In the case of Saturn, its geocentric right ascension $\alpha$, declination $\delta$ and range $\rho$ would be affected up to $10^4-10^5$ milliarcseconds and $10^5$ km, respectively; the present-day residuals of such observables are as little as about $4$ milliarcseconds and $10^{-1}$ km, respectively. \textcolor{black}{We analytically calculated the long-term orbital effects induced by the putative harmonic variation of $G$ at hand finding non-zero rates of change for the semimajor axis $a$, the eccentricity $e$ and the argument of pericenter $\omega$ of a test particle. For the LAGEOS satellite, an orbital increase as large as $3.9$ m yr$^{-1}$ is predicted, in contrast with the observed decay of $-0.203\pm 0.035$ m yr$^{-1}$. An anomalous perihelion precession as large as 14 arcseconds per century is implied for Saturn, while latest observations constrain it down to the $10^{-4}$ arcseconds per century level. \textcolor{black}{The rejection level provided by Mercury is of the same order of magnitude.}}
\end{abstract}


%


\section{Introduction}
The Newton's gravitational constant $G$ \citep{1997RPPh...60..151G,2012RvMP...84.1527M}, measured for the first time\footnote{For a critical discussion of earlier attempts \citep{bou,mask}, see, e.g., \citet{poy,sma}.} by \citet{cav} at the end of the eighteenth century\footnote{In view  of the unit conventions  in use at the time of Cavendish's work, the gravitational constant did not appear explicitly in it. Indeed, one of the first explicit formulations of the Newtonian gravity in terms of a gravitational constant $G$ appeared not before 1873 \citep{prima}. Thus, the English scientist actually measured the mean density of the Earth \citep{1987AmJPh..55..210C}. His results were later reformulated in modern terms as a determination of $G$ \citep{poy,mac,boys}. }, is one of the fundamental parameters of Nature setting the magnitude of the gravitational interaction \citep{uzan1,uzan2,uzan3,2011PThPh.126..993C}. In either the Newtonian and the Einsteinian theories it is assumed that it does depend neither on spatial nor temporal coordinates, being a truly universal constant.

In the twentieth century, prominent scientists \citep{1935rgws.book.....M,1937RSPSA.158..324M,1937Natur.139..323D,1937NW.....25..513J,1939ZPhy..113..660J}, mainly on the basis of cosmological arguments, argued that, actually, $G$ may experience slow time variations over the eons.
Current research took over such a fascinating idea, so that nowadays there are several theoretical scenarios encompassing it; see, e.g., \citet{1961PhRv..124..925B,1962PhRv..125.2194B,1986PhRvL..57.1978W,1988NCimB.102..131I,2002IJMPA..17.4325M,2009FrPhC...4...75M}.

Recently, \citet{gsine} showed that \textcolor{black}{a set of} measurements of $G$ obtained \textcolor{black}{over the years} with different techniques (see, e.g., \citep{2000PhRvL..85.2869G,2001PhRvL..87k1101Q,2006PhRvD..74h2001S,2007Sci...315...74F,2008PhRvL.100e0801L,2010PhRvL.105k0801P,2009PhRvL.102x0801L,2010PhRvD..82b2001T,2013PhRvL.111j1102Q,2014Natur.510..518R,2014Natur.510..478S}) in terrestrial laboratories over the latest decades \citep{2014PhT....67g..27S} \textcolor{black}{can be satisfactorily modeled with the following harmonic time-dependent signature}
\eqi G(t) = G_0 + \Delta G(t)= G_0 + A_G\sin\ton{\omega_G t + \phi},\lb{gitti}\eqf with \citep{gsine}
\begin{align}
G_0 &= 6.673899\times 10^{-11}~\textrm{kg}^{-1}~\textrm{m}^3~\textrm{s}^{-2}, \\ \nonumber \\
A_G \lb{aggi} &=1.619\times 10^{-14}~\textrm{kg}^{-1}~\textrm{m}^3~\textrm{s}^{-2},\\ \nonumber \\
\phi & = 80.9~\textrm{deg}, \\ \nonumber \\
\omega_G &=\rp{2\pi}{P_G},\\ \nonumber \\
P_G \lb{piggi}&=5.899~\textrm{yr}.
\end{align}
\textcolor{black}{\citet{gsine} did not suggest that the reported pattern may be due to some modifications of the currently accepted laws of gravity. They remarked that a correlation with recently reported harmonic variations in measurements of the Length of Day (LOD) \citep{holme} is present. Later,
 a compilation of all published measurements of $G$ performed since 1980, including also some additions and corrections with respect to the values analyzed in \citet{gsine}, was offered by \citet{2015PhRvD..91l1101S}. Their least-square fit to such an expanded dataset confirmed the existence of the sinusoidal component with a $5.9$ yr periodicity by \citet{gsine}, although the correlation with the LOD turned out to be weakened; in the minimization procedure, a second periodicity of about 1 yr was found by \citet{2015PhRvD..91l1101S}. Cautiously, \citet{2015PhRvD..91l1101S} did not explicitly mention any possible variation of the gravitational laws, warning that the harmonic pattern they found may be due to some-unspecified-systematic errors, with underestimated systematic uncertainty.
\citet{2015EL....11130002P}, by using the compiled measurements of $G$  by \citet{2015PhRvD..91l1101S} and a Bayesian  model comparison, claimed that a constant $G$ measurement model with an additional Gaussian noise term would fit the experimental data better than a model containing periodic terms. In the reply they offered, \citet{2015EL....11130003A} were not able to independently confirm the claim by \citet{2015EL....11130002P}, and stood by their conclusions of potential periodic terms in the reported $G$ measurements. In particular, \citet{2015EL....11130002P} found that a model with two harmonic components would be slightly favored with respect to their early proposal \citep{gsine}, allowing also to cope with the issue of the weakened correlation with the LOD  pointed out in \citet{2015PhRvD..91l1101S}.
}

\textcolor{black}{At the time of writing, none of the putative systematic errors allegedly affecting the Earth-based measurements of $G$, reasonably conjectured  by \citet{gsine} and \citet{2015PhRvD..91l1101S}, have been yet disclosed\footnote{\textcolor{black}{A search for them is outside the scope of the present work.}}. \textit{Things standing thus unknown}, it is not unreasonable to follow a complementary approach, which could help in effectively selecting the directions of further experimental analyses, and look at the potentially intriguing-although admittedly unlikely-possibility that some modifications of the currently accepted laws of gravity may be at work in the present case. According to \citet{2015arXiv150407622K}, the observed discrepancies between the $G$ values determined by different experiments may be connected with a differential interpretation of MOND theory applied to the galaxy rotation curves.}

\textcolor{black}{In this spirit, we want to quantitatively put to the test the daring conjecture that (some of) the observed sinusoidal pattern(s) in the $G$ data records may reflect an unexpected physical phenomenon} in an independent way by looking at the consequences that such an effect, if real, would have on systems other than those used to collect the measured values of $G$ on the Earth. To this aim, \textcolor{black}{for the sake of simplicity,} we will consider  \textcolor{black}{the simpler model of \rfr{gitti}-\rfr{piggi} and} the changes which would occur in the motions of the major bodies of the Solar System to check if they are compatible with the current stringent limits posed on their standard dynamics by accurate astrometric measurements.

Here, we will use recently released Cassini data analyses spanning ten years (2004-2014) of the orbit of Saturn in terms of its geocentric range $\rho$, right ascension $\alpha$, declination $\delta$ \citep{2014PhRvD..89j2002H,2014arXiv1410.1067J} to  independently test the $G(t)$ scenario of
\rfr{gitti}-\rfr{piggi}. In particular, we will suitably compare numerically simulated signatures $\Delta\alpha (t),~\Delta\delta (t),~\Delta\rho (t)$ induced by \rfr{gitti}-\rfr{piggi} on the Kronian Celestial coordinates with the currently existing residuals for them \citep{2014PhRvD..89j2002H,2014arXiv1410.1067J}. Full details of the methodology adopted are given in Section \ref{metodi}. Section \ref{theend} summarizes our findings.
\section{Methods and results}\lb{metodi}
A striking feature of the alleged time-variation of $G$ investigated here is its relatively short characteristic time scale which, according to \rfr{piggi}, amounts to just about 6 yr. This is in neat contrast with virtually all the theoretical models predicting a $G(t)$ varying over typically cosmological timescales. This distinctive feature has also direct phenomenological consequences. Indeed, the validity of the numerous bounds on the percent variation of $G$ existing in the literature, of the order of \citep{2004PhRvL..93z1101W,2007CQGra..24r4533M,2013AstL...39..141P,2013MNRAS.432.3431P} \eqi\left|\rp{\dot G}{G}\right|\sim 10^{-13}-10^{-14}~\textrm{yr}^{-1},\eqf may not be straightforwardly extended to the present case since they were inferred from least-square reductions of planetary and lunar positional data by modeling $\Delta G(t)$ as a secular trend. Such a choice, reasonable in view of the extremely slow changes assumed in the literature for $G$ with respect to the typical orbital frequencies of the major bodies of the Solar System, does not apply to \rfr{gitti}. Thus, a dedicated analysis should be performed in the present case: it will be the subject of the present Section. \textcolor{black}{In particular, in Section \ref{numero} a numerical approach will be followed, while an analytical calculation will be offered in Section \ref{anale}.}
\subsection{A numerical approach}\lb{numero}
As a first step, we simultaneously integrate the barycentric equations of motion of all of the currently known major bodies of the Solar system in rectangular Cartesian coordinates over a centennial time-span (1914-2014) with the standard package MATHEMATICA\footnote{The MATHEMATICA method adopted is {ExplicitRungeKutta}, with a working precision used in internal computations of $53 \log_{10}2 \approx 16$, and 8 digits of precision and of absolute accuracy. The computer used has a 64-bit operating system.}. \textcolor{black}{The dynamical accelerations modeled include the General Theory of Relativity to the first Post-Newtonian level, and all the major known Newtonian effects like the Sun's oblateness, pointlike mutual perturbations by the eight planets and the three largest asteroids, two massive rings accounting for the minor asteroids \citep{2010A&A...514A..96K} and the Kuiper Belt's objects \citep{2013MNRAS.432.3431P}. } The initial conditions are taken from Tables \ref{xyz} to \ref{vxvyvz}:
\begin{table*}

\centering
\caption{Solar System Barycentric (SSB) initial positions $x_0,~y_0,~z_0$ of the Sun, the eight planets  and the dwarf planet Pluto \textcolor{black}{estimated} with the EPM2013 ephemerides \citep{Pit014}. They are referred to the International Celestial Reference Frame (ICRF2) at the epoch JD 2446000.5 (27 October 1984, h: 00.00.00). (E.V. Pitjeva, private communication).}
\label{xyz}
\begin{tabular}{llll}
\noalign{\smallskip}
\hline
 & $x_0$ $\ton{\textrm{au}}$ & $y_0$ $\ton{\textrm{au}}$ & $z_0$ $\ton{\textrm{au}}$ \\
\hline
Sun & $ 0.0005203216237770 $ & $ 0.0084630932909853 $ & $ 0.0035304787263118 $\\
Mercury & $ -0.1804160787007675 $ & $ -0.3766844130175976 $ & $ -0.1834161566766066 $\\
Venus & $ 0.3542041788200310 $ & $ -0.5632690218672094 $ & $ -0.2760410432968001 $\\
Earth & $ 0.8246466037849634 $ & $ 0.5178999919107230 $ & $ 0.2244221978752543 $\\
Mars & $ 1.1810208746662094 $ & $ -0.6327726058206141 $ & $ -0.3225338798256297 $\\
Jupiter & $ 1.6184148186514691 $ & $ -4.4994366329965958 $ & $ -1.9681878191442090 $\\
Saturn & $ -6.5110893457860746 $ & $ -6.9749292526618341 $ & $ -2.6006192842623910 $\\
Uranus &  $-5.3939091822416660$ &   $-16.7458210494500221$  &   $-7.2578236561504630$ \\
Neptune &  $0.5400549266512163$  &  $-27.9860579257785673$  &  $-11.4683596423229197$ \\
Pluto & $-24.1120716069166541$  &  $-17.4030370013573190$   &   $1.8337998706332588$ \\
\hline
\end{tabular}
\centering
\caption{SSB initial velocities $\dot x_0,~\dot y_0,~\dot z_0$ of the major bodies of the Solar system. The other details are as in Table \ref{xyz}.}
\label{vxvyvz}
\begin{tabular}{llll}
\noalign{\smallskip}
\hline
 & $\dot x_0$ $\ton{\textrm{au~d}^{-1}}$ & $\dot y_0$ $\ton{\textrm{au~d}^{-1}}$ & $\dot z_0$ $\ton{\textrm{au~d}^{-1}}$ \\
\hline
Sun & $  -0.0000082057731502 $ & $  -0.0000014976919650 $ & $ -0.0000004352388239 $\\
Mercury & $ 0.0202454878200986 $ & $ -0.0077363989584885 $ & $ -0.0062333734254165 $\\
Venus & $ 0.0175313054842932 $ & $ 0.0093063903885975 $ & $ 0.0030763136627655 $\\
Earth & $ -0.0099025409715828 $ & $ 0.0130300000107108 $ & $ 0.0056499201171062 $\\
Mars & $ 0.0078002404125410 $ & $ 0.0120303517832832 $ & $ 0.0053067080995448 $\\
Jupiter & $ 0.0070681002624717 $ & $ 0.0025516512292944 $ & $ 0.0009215336853663 $\\
Saturn & $ 0.0038982964656769 $ & $ -0.0033491220519229 $ & $ -0.0015508256594232 $\\
Uranus &  $0.0037423336376372$  &   $-0.0011699461011685$  &   $-0.0005654073338528$ \\
Neptune &  $0.0031184897780294$  &    $0.0000964171123061$   &  $-0.0000381527005270$\\
Pluto & $0.0019439712118417$  &   $-0.0025828344901677$  &   $-0.0013916896197374$\\
\hline
\end{tabular}
\end{table*}
they come from an adjustment  of the suite of measurement and dynamical models of the EPM2013 ephemerides \citep{Pit014} to an extended data record of more than $800\,000$ observations ranging covering last century, and are referred to the epoch $t_0 = \textrm{JD}~2446000.5$ (27 October 1984, h: 00.00.00).

\begin{figure*}
\centering
\centerline{
\vbox{
\begin{tabular}{cc}
\epsfysize= 5.0 cm\epsfbox{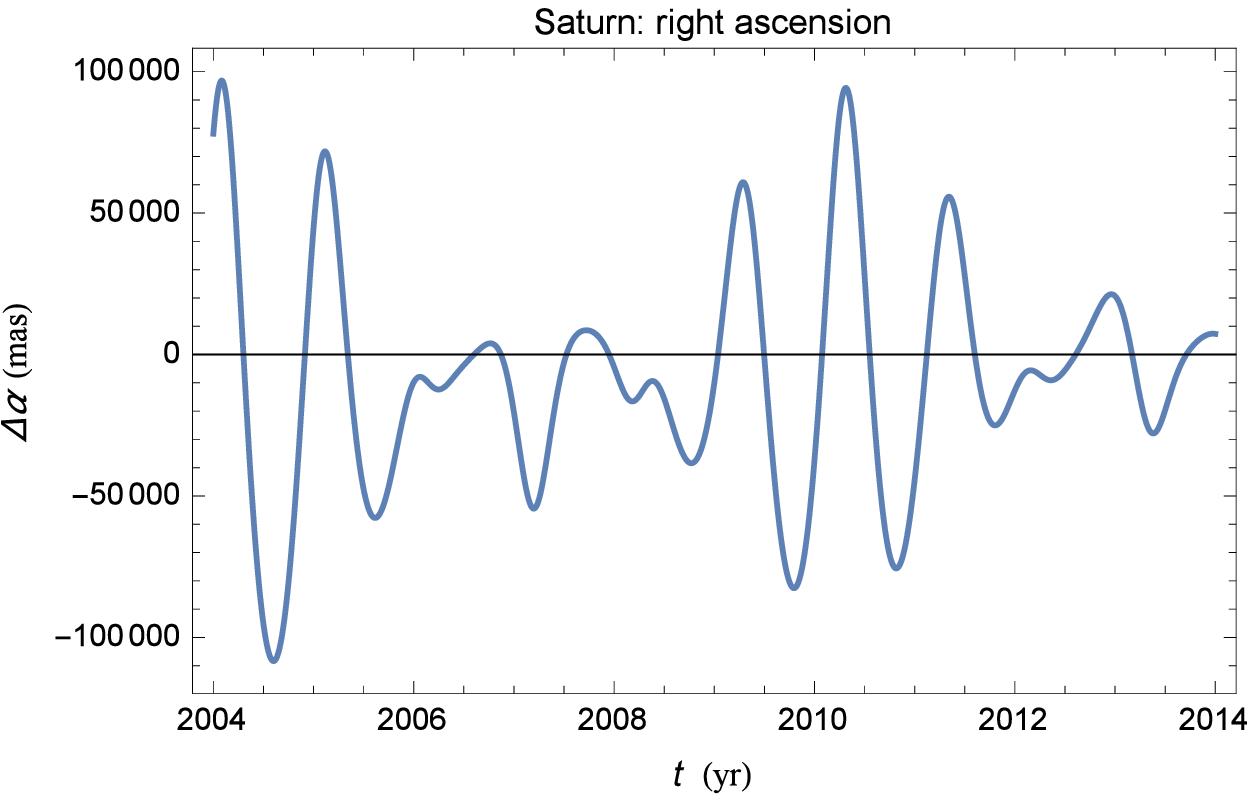} \\
\epsfysize= 5.0 cm\epsfbox{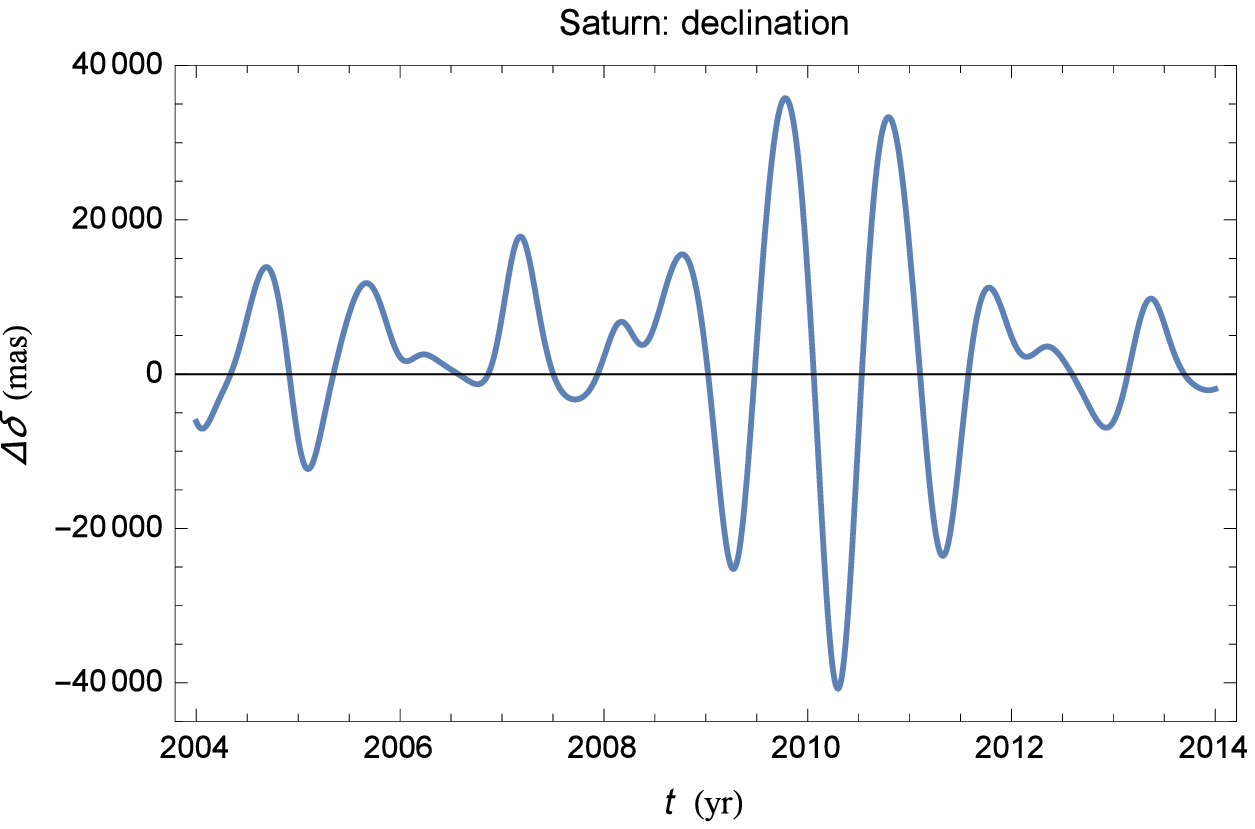} \\
\epsfysize= 5.0 cm\epsfbox{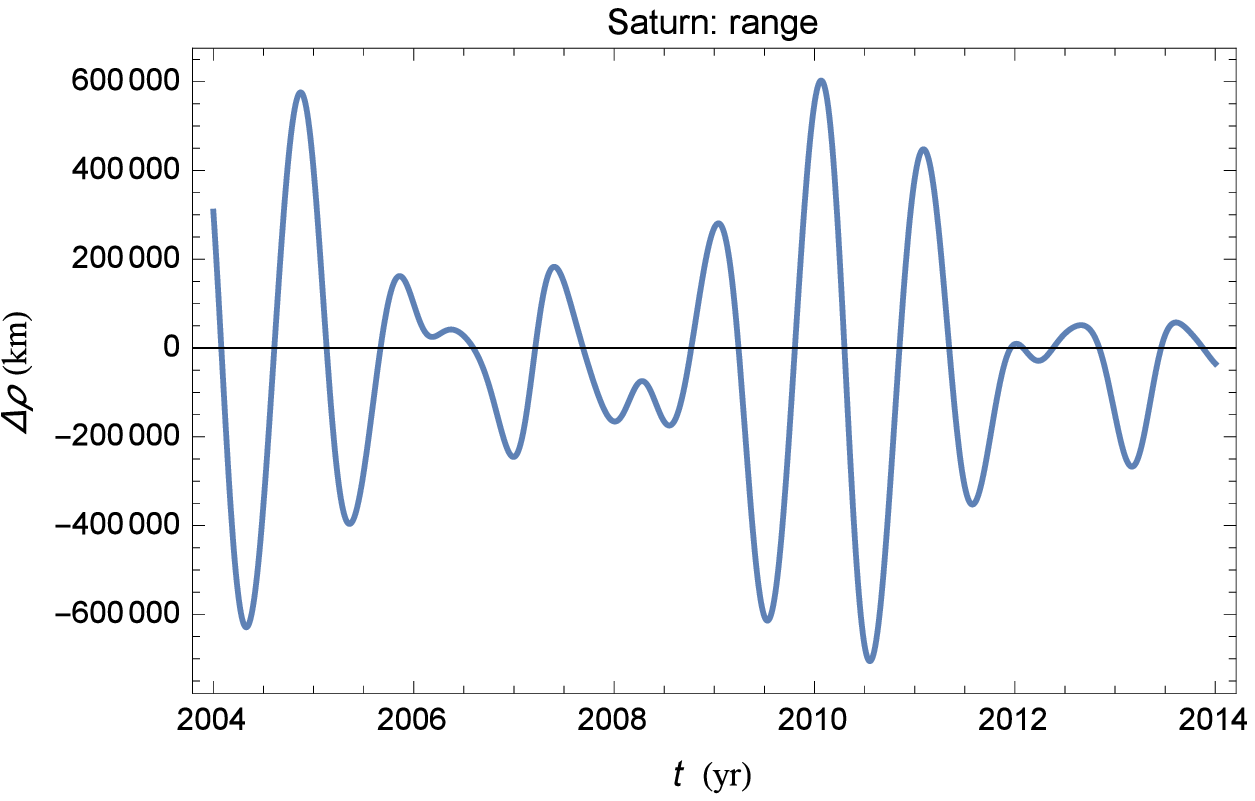} \\
\end{tabular}
}
}
\caption{Numerically produced time series of $\rho,~\alpha,~\delta$ of Saturn perturbed by $\Delta G(t)$ as in \rfr{gitti}-\rfr{piggi}. For each of the three Kronian observables considered, they were calculated as differences between two numerical integrations of the SSB barycentric equations of motion of the Sun, its eight planets and the dwarf planet Pluto from 1914 to 2014 with and without $\Delta G(t)$. Both integrations shared the same initial conditions in rectangular Cartesian coordinates, retrieved from Tables \ref{xyz} to \ref{vxvyvz}, and the same standard dynamical models, apart from $\Delta G(t)$ itself.
Thus, such curves represent the expected $\Delta G(t)$-induced signatures  $\Delta\rho,~\Delta\alpha,~\Delta\delta$.
The patterns and the size of the present signals can be compared with the range residuals by \citet{2014PhRvD..89j2002H} ($\Delta\rho_{\rm exp}\lesssim 0.1$ km) and those for RA, DEC by \citet{2014arXiv1410.1067J} ($\Delta\alpha_{\rm exp},\Delta\delta_{\rm exp}\lesssim 4$ milliarcseconds).}\label{figura}
\end{figure*}

%
Thus, keeping the other parameters of the numerical integration unchanged, we repeat the same step  by including also the putative variation of $G$ according to \rfr{gitti}-\rfr{piggi}. Both  numerical integrations, with and without $\Delta G(t)$, share the same initial conditions for the known bodies of the Solar system retrieved from Tables \ref{xyz} to \ref{vxvyvz}. From the resulting time series of the Earth and Saturn,
we numerically compute the time series of $\rho,~\alpha,~\delta$ of Saturn, with and without $\Delta G(t)$; then, for all of the three Kronian Celestial coordinates, we compute differential time series $\Delta\alpha (t),~\Delta\delta (t),~\Delta\rho (t)$  which show up the impact of $\Delta G(t)$ over the 2004-2014 interval of time covered by the most recent Cassini data analyses \citep{2014PhRvD..89j2002H,2014arXiv1410.1067J}; also Gaussian white noise is added to properly simulate the impact of the measurement errors.
The results are displayed in Fig. \ref{figura}.
Finally, we compare our simulated residuals to the corresponding existing Kronian residuals in Fig. 5 of \citet{2014PhRvD..89j2002H} and Fig. 2 of \citet{2014arXiv1410.1067J}, which were produced without explicitly modelling the perturbing action of $\Delta G(t)$.

Some technical considerations about the general validity of the approach followed are in order.
Possible objections of lacking of meaningfulness concerning such kind of direct comparisons among theoretically calculated signatures of a certain dynamical effect and actual data processed without modelling the effect itself have recently proved to be ineffective, at least in some specific cases. Indeed, apart from the fact that such an approach had been proven successful since the time of the Pioneer anomaly \citep{2006NewA...11..600I, 2008AIPC..977..254S, 2010IAUS..261..179S, 2010IAUS..261..159F}, the latest constraints on a certain form of the MOND theory, equivalent to the action of a remote trans-Plutonian body located in the direction of the Galactic Center, which were obtained by explicitly model{l}ing it in a dedicated planetary data reduction \citep{2014PhRvD..89j2002H}, turned out to be equivalent to those previously established by comparing theoretically computed effects to their observationally inferred counterparts determined without modelling it \citep{2010OAJ.....3....1I}.
Finally, one might raise consistency issues about our analysis since our initial conditions come from the EPM2013 ephemerides \citep{Pit014}, while the post-fit residuals \citep{2014PhRvD..89j2002H,2014arXiv1410.1067J} to which our simulated signatures are contrasted were obtained with the DE430 ephemerides. Actually, it is not so.
First, the EPM2013 initial planetary state vectors differ from the  DE431 coordinates at $t_0$, retrieved from the HORIZONS Web interface at http://ssd.jpl.nasa.gov/?horizons, by just $\epsilon=2\left(x_{\rm EPM}-x_{\rm DE}\right)/\left(x_{\rm EPM}+x_{\rm DE}\right)\approx 10^{-9}$. Then, from Fig. 7 of \citet{2013SoSyR..47..386P} it can be noticed that the differences between the Kronian Celestial coordinates calculated with the EPM2011 and DE424 ephemerides over a time interval as little as 10 years (2004-2014) are smaller than the residuals in \citet{2014PhRvD..89j2002H,2014arXiv1410.1067J} obtained with the DE430 ephemerides.
\subsection{\textcolor{black}{An analytical calculation}}\lb{anale}
\textcolor{black}{
The sinusoidal part of $G$ induces a time-dependent component of the gravitational acceleration which, to the Newtonian level, can be written as
\eqi A_G = -\rp{\Delta G(t) M}{r^2}.\lb{pertu}\eqf In view of \rfr{aggi}, \rfr{pertu}  can be thought as a small radial correction to the usual inverse-square law. As such, it can be treated with the standard methods of the perturbation theory \citep{2003ASSL..293.....B, 2008orbi.book.....X} by assuming a Keplerian ellipse as unperturbed, reference trajectory.
}

\textcolor{black}{
For a test particle orbiting a primary of mass $M$ along a Keplerian ellipse with semimajor axis $a_0$, orbital period $\Pb  = 2\pi n_0^{-1}=2\pi\sqrt{a_0^3 G_0^{-1} M^{-1}}$ and eccentricity $e_0$, the time $t$ is connected with the true anomaly $f$, which gives the instantaneous position along the orbit, through \citep{Capde05}
\eqi n_0\ton{t-t_p} = 2\arctan\qua{\sqrt{\rp{1-e_0}{1+e_0}}\tan\ton{\rp{f}{2}}} -\rp{e_0\sqrt{1-e_0^2}\sin f}{1+e_0\cos f}.\lb{tempo}\eqf
In \rfr{tempo},  $n_0$ is the unperturbed Keplerian mean motion, while $t_p$ is the time of passage at pericenter. The true anomaly $f$ usually starts to be reckoned just at the crossing of the pericenter, so that $f\ton{t_p}=0$. For the sake of simplicity, in the following we will assume $t_p = 0$.
}

\textcolor{black}{
By inserting \rfr{pertu}, with \rfr{gitti} and $t$ given explicitly by \rfr{tempo}, in the right-hand sides of the Gauss equations \citep{2003ASSL..293.....B, 2008orbi.book.....X} for the variation of the Keplerian orbital elements\footnote{In \rfr{dAdt}-\rfr{dperidt}, $A_R,~A_T,~A_N$ are the components of the disturbing acceleration $\bds A$ onto the radial, transverse and out-of-plane directions, $\Omega$ is the longitude of the ascending node, $\omega$ is the argument of pericenter, and $I$ is the orbital inclination to the reference $\grf{x,y}$ plane.}
\begin{align}
\dert a t \lb{dAdt}& = \rp{2}{n_0\sqrt{1-e_0^2}}\qua{e_0 A_R\sin f + A_T\ton{\rp{p_0}{r_0}} }, \\ \nonumber \\
\dert e t & = \rp{\sqrt{1-e_0^2}}{n_0 a_0}\grf{A_R\sin f + A_T\qua{\cos f + \rp{1}{e_0}\ton{1-\rp{r_0}{a_0}}  } }, \\ \nonumber \\
\dert\Omega t & = \rp{1}{n_0 a_0\sqrt{1-e_0^2}\sin I_0}A_N\ton{\rp{r_0}{a_0}}\sin\ton{\omega_0 + f}, \\ \nonumber \\
\dert\omega t \lb{dperidt}&= \rp{\sqrt{1-e_0^2}}{n_0 a_0 e_0}\qua{-A_R\cos f + \ton{1 + \rp{r_0}{p_0}}\sin f} - \cos I_0\dert\Omega t,
\end{align}
 evaluated onto the unperturbed Keplerian ellipse
\eqi r_0 = \rp{p_0}{1+e_0 \cos f},~p_0\doteq a_0\ton{1-e_0^2},\eqf
and averaging them over one orbital period $\Pb$ by means of \citep{Capde05}
 \eqi
 \dert t f = \rp{\ton{1-e_0^2}^{3/2}}{n_0\ton{1 + e_0\cos f}^2},
 \eqf
 non-vanishing long-term rates of change of the semimajor axis, the eccentricity and the pericenter  are obtained. They are
\begin{align}
\ang{\dert a t} \lb{dadt} & = -\rp{e_0 A_G M\Pb P^2_G}{\pi^2 a_0^2\ton{\Pb^2 - P_G^2}}\cos\ton{\rp{\pi\Pb}{P_G}+\phi}\sin\ton{\rp{\pi\Pb}{P_G}}  + \mathcal{O}\ton{e_0^2},\\ \nonumber \\
\ang{\dert e t} & = \rp{A_G M\Pb P^2_G}{4\pi^2 a_0^3\ton{\Pb^2 - P_G^2}}\qua{\sin\phi - \sin\ton{\rp{2\pi\Pb}{P_G} + \phi} } + \mathcal{O}\ton{e_0},\\ \nonumber \\
\ang{\dert \omega t} \lb{dodt} & = \rp{A_G M\Pb^2 P_G}{\pi^2 a_0^3\ton{\Pb^2 - 4P_G^2}}\sin\ton{\rp{\pi\Pb}{P_G}}\sin\ton{\rp{\pi\Pb}{P_G}+\phi} + \mathcal{O}\ton{e_0}.
\end{align}
While the leading component of the rate of change of the semimajor axis is of order $\mathcal{O}\ton{e_0}$, the next-to-leading order terms of the rates of the eccentricity and the pericenter are of order $\mathcal{O}\ton{e_0}$.
}

\textcolor{black}{
To the approximation level in $e$ indicated, which is quite adequate in the Solar System,  \rfr{dadt}-\rfr{dodt} are exact in the sense that no a priori assumptions on the relative magnitudes of $\Pb$ and $P_G$ were taken. Thus, they can be applied to a variety of systems ranging, e.g., from the fast Earth's artificial satellites to the slowest planet of the Sun. As such, \rfr{dadt}-\rfr{dodt}  hold also for any putative modified model of gravity yielding possibly a harmonic variation of $G$ over arbitrary timescales \textcolor{black}{\citep{1990ApJ...362L..37M, 1993PhRvD..47.5329B, 1993PhRvD..48.3592B, 1994PhRvD..50.3746B, 1995PhRvD..51.2729B, 1996PhRvD..54.3920B, 1997PhRvD..55.1906B, 2015PhRvL.115t1301S}. }
}

Let us start with the LAGEOS satellite orbiting the Earth along a nearly circular orbit with $a_0 = 12274~\textrm{km},~e_0=0.0039,~\Pb = 3.7~\textrm{hr}=4.2\times 10^{-4}~\textrm{yr}$. It is known since long time \citep{1982CeMec..26..361R} that its semimajor axis experiences a secular decrease due to a variety of physical mechanisms. Its latest measurement amounts to \citep{Sosproc, SosBOOK}
\eqi\dot a^{\rm meas}_{\rm LAGEOS}  = -0.203\pm 0.035~\textrm{m~yr}^{-1}. \lb{soss}\eqf
Instead, \rfr{dadt} predicts a secular increase as large as
\eqi\dot a^{\rm pred}_{\rm LAGEOS} =3.9~\textrm{m~yr}^{-1},\eqf which is in disagreement with \rfr{soss}.

Moving to Saturn, which orbits the Sun in about 29 yr at $9.5$ au, it turns out that \rfr{dadt} and \rfr{dodt} predict
\begin{align}
\dot a^{\rm pred}_{\saturn} & = -668~\textrm{m~yr}^{-1}, \\ \nonumber \\
\dot\omega^{\rm pred}_{\saturn} & = 14~ '' ~\textrm{cty}^{-1}.
\end{align}
Actually, there is no trace of such an enormous secular decrease of the Kronian semimajor axis in the observational data records. As far as its perihelion is concerned, the current admitted range for a putative anomalous precession is as little as\footnote{Here, $\varpi\doteq\Omega+\omega$ is the longitude of pericenter.} \citep{2013MNRAS.432.3431P, 2011CeMDA.111..363F}
\begin{align}
\Delta\dot\varpi^{\rm EPM}_{\saturn} &= \ton{-3.2\pm 4.7}\times 10^{-4}~ '' ~ \textrm{cty}^{-1}, \\ \nonumber \\
\Delta\dot\varpi^{\rm INPOP}_{\saturn} &= \ton{1.5\pm 6.5}\times 10^{-4}~ '' ~ \textrm{cty}^{-1}.
\end{align}
\textcolor{black}{Similar results hold also for other planets like, e.g., Mercury. His predicted perihelion rate is as large as
\eqi
\dot\omega^{\rm pred}_{\mercury}  = -108~ '' ~\textrm{cty}^{-1}, \eqf while any anomalous precession is constrained within
\citep{2013MNRAS.432.3431P, 2011CeMDA.111..363F}
\begin{align}
\Delta\dot\varpi^{\rm EPM}_{\mercury} &= \ton{-2\pm 3}\times 10^{-3}~ '' ~ \textrm{cty}^{-1}, \\ \nonumber \\
\Delta\dot\varpi^{\rm INPOP}_{\mercury} &= \ton{4\pm 6}\times 10^{-4}~ '' ~ \textrm{cty}^{-1}.
\end{align}
 }
\section{Summary and conclusions}\lb{theend}
In this paper, we have independently tested the hypothesis that the  harmonic temporal variation in the time series of the laboratory measurements of the Newtonian constant of gravitation made over the years, which has been recently reported in the literature, may be due to some unknown physical mechanism affecting $G$ itself. Cautiously, its discoverers did not mention such a possibility, claiming instead that it may be likely due to some systematic errors. Nonetheless, at present, none of them has been either identified or even explicitly suggested.

To this aim, we looked at the effects that such a putative time-dependent behaviour of the fundamental parameter characterizing the strength of the gravitational interaction, if real, would have on the orbital motions of the major bodies of the Solar System. We numerically integrated their equations of motion with and without the proposed modification, and calculated the differences of the resulting time series for the observables used in real astrometric data reductions (right ascension $\alpha$, declination $\delta$, range $\rho$) to produce simulated residual signals $\Delta\alpha (t),~\Delta\delta (t),~\Delta\rho (t)$.
We remark that, given the specific functional dependence of the effect considered and its relatively short characteristic time scale compared to the typical Solar System's orbital periods, it would be incorrect to straightforwardly extend the existing bounds on $\dot G/G$ to the present case since they were obtained by modeling a secular variation of $G$.
It turned out that the resulting anomalous signatures in right ascension, declination and range are far too large to have escaped detection in the residuals produced so far with the existing standard ephemerides, even if the putative variation of $G$ was not explicitly modeled in all of them.
Suffice it to say that, in the case of Saturn, $\alpha$, $\delta$ and $\rho$ would be affected up to $10^4-10^5$ milliarcseconds and $10^5$ km, while the current residuals are as little as about $4$ milliarcseconds and $10^{-1}$ km, respectively.

Complementarily, we performed also an analytical calculation of the long-term perturbations which a harmonic variation of $G$ would induce  on the orbital elements of a test particle orbiting a central mass. In the limit of small eccentricities, we obtained non-vanishing secular variations of the semimajor axis $a$, the eccentricity $e$ and the pericenter $\omega$. They have a general validity since no a priori assumptions concerning the relative magnitudes of the orbital and $G$ frequencies  were made\textcolor{black}{; as such, they can be applied, in principle, to any physical mechanism predicting a sinusoidally time-dependent variation  of $G$.}  A comparison with the latest observational determinations for the LAGEOS satellite and Saturn confirmed the outcome of the numerical analysis. Indeed, the recently measured orbital decay of the geodetic satellite, of the order of $-0.2\pm 0.03$ m yr$^{-1}$, rules out the predicted increase of $\approx +4$ m yr$^{-1}$. Furthermore, the expected anomalous apsidal rate of 14 arcseconds per century for Saturn falls neatly outside of the allowed range for any possible anomalous Kronian perihelion precession, which is set by the observations to the $\approx 10^{-4}$ $''$ cty$^{-1}$ level. \textcolor{black}{Also Mercury yields an analogous outcome since the magnitude of its predicted anomalous perihelion precession amounts to 108 $''$ cty$^{-1}$, while the observations constrain any possible deviations from standard physics  down to $\approx 10^{-3}-10^{-4}$ $''$ cty$^{-1}$.}

In conclusion, our analysis quantitatively rules out the possibility that some long-range modification of the currently accepted laws of the gravitational interaction may be at work in the present case accounting for the observed harmonic pattern of the laboratory-measured values of $G$. As such, it may contribute to further direct future investigations towards the discovery of the even more likely systematic uncertainties allegedly plaguing the set of measurements of $G$ analyzed  so far.
\bibliography{Gsinebib}{}

\end{document}